# Kinetic effects in diffusion on a disordered square lattice

R.J. Dikken[1], C.D. Versteylen[1,2]

[1]Department of Materials Science and Engineering, Delft University of Technology, 2628CD Delft, the Netherlands

[2]Faculty of Applied Sciences, Delft University of Technology, 2629 JB Delft, The Netherlands

**Abstract**

In this work, the effect of fluctuations in a disordered square lattice on diffusion of a test particle is studied using kinetic Monte Carlo simulations. Diffusion is relevant to a wide variety of problems, both within physics and outside of physics. Kinetic effects in diffusion are often hidden in a thermodynamical description of the problem. In this work, no assumptions based on energy are made, and diffusion occurs purely based on the attempt rate of the test particle and the occupation and fluctuation rate of the lattice. Although the average transition rate of the particle is the same for a static or fluctuating lattice with specific occupation, the diffusion constant is kinetically affected in a fluctuating, disordered lattice. If the lattice fluctuates faster than the attempt rate of the particle, diffusion is controlled by the attempt rate of the particle. However, if the lattice fluctuates slower than the attempt rate of the particle, diffusion is affected by the fluctuations. The slower the lattice fluctuates, the lower the diffusion constant. Furthermore, it is found that for fast fluctuating lattices, diffusion is due to Brownian motion. If the lattice fluctuates slower than the particle, diffusion becomes anomalous depending on the occupation of the lattice.

## I Introduction

Diffusion applies to a large variety of problems. Diffusion related problems in physics involve well known problems like heat transport, diffusion of lubricants in rough contacts, percolation, diffusion in solids [1] etc. However, also outside of physics, diffusion is pivotal in biological and socio-technical systems and involves for instance epidemic spreading [2, 3].

Although the effect of fluctuations of a disordered environment on diffusion is intensively studied over the past decades a full understanding is still not reached [4]. This is partly attributed to the combined effects of thermodynamics and kinetics. Especially when different time scales apply to a disordered environment and intrinsic to the particle that diffuses, it is not trivial to determine what controls the diffusion constant. The aim of this study is to gain understanding of the kinetic effects of a fluctuating, disordered lattice on the diffusion of a test particle.

Because of its simple geometry, a two-dimensional square lattice is especially convenient for studying diffusion of a test particle in a fluctuating, disordered environment. In a static lattice, so without fluctuations in the site occupation, the diffusion involves a percolation problem [5, 6]. It is known that fluctuations in the occupation of a lattice can affect the diffusion behavior of a test particle, when the sites to which the particle can move fluctuate according to a Poisson process, introducing

Corresponding author: R.J. Dikken, robbertjandikken@hotmail.com

a memory to the lattice sites [7]. The memory of the lattice sites makes that the diffusion behavior is non-Gaussian. If the fluctuation time approaches infinity, the problem involves a percolation transition as found for several lattices geometries [8].

That kinetic effects are involved in diffusion is known for more than a century [9,10]. However, as mentioned, thermodynamics and kinetics are often combined, since in many problems a certain chemical potential is involved. In this work, the formulation does not involve any energy, which effectively means that all states are energetically equivalent, so that we can study the pure effect of kinetics of the environment on the diffusion of a particle. To this end, we focus on a disordered, two-dimensional square lattice for which the sites are occupied with a certain probability. A test particle attempts a transition with a certain rate, but can only move to an unoccupied site. The lattice fluctuates randomly with a certain rate. By tracking the particle and calculating the mean squared displacement the diffusion constant can be obtained. Also it is investigated where diffusion deviates from Brownian motion [9] and becomes anomalous [4].

## II  Problem formulation

In this study the diffusion of a test particle on a two-dimensional square lattice (periodic in both directions) with random disorder is studied using a formulation in which all states are energetically equivalent. The lattice consists of sites that can either be occupied with a probability $f$ or empty with a probability $1 - f$. No interaction between the sites is assumed. The test particle can only move to nearest neighbor sites that are unoccupied. Figure 1 shows an example of a lattice. The green dot represents the test particle and the blue open and solid dots represent open and occupied lattice sites, respectively. Since random motion is assumed

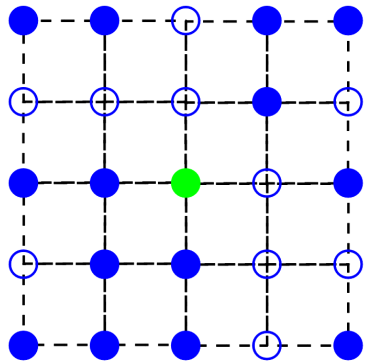

Figure 1: Lattice example with $f = 0.56$.

for the test particle, the mean transition rate is given by

$$\Gamma_{\text{tr}} = (1 - f)\Gamma_{\text{p}} \quad (1)$$

where $\Gamma_{\text{p}}$ is the attempt rate of the test particle. In the calculations, time is normalized on the fastest rate and space is normalized such that one jump is of unit length. The mean square displacement $d(t)$ given by

$$d(t) = \langle (\mathbf{r}(t_0 + t) - \mathbf{r}(t_0))^2 \rangle \quad (2)$$

where $\mathbf{r}$ is the displacement in two dimensions. The mean square displacement is proportional with time:

$$d(t) \propto t^\alpha. \quad (3)$$

For $\alpha \neq 1$, diffusion is anomalous (superlinear for $\alpha > 1$ and sublinear for $\alpha < 1$). For $\alpha = 1$ diffusion is according to Brownian motion. The proportionality constant between mean square displacement and time is then the diffusion constant. So the diffusion constant is given by [9,11]:

$$D = \lim_{t \to +\infty} \frac{d(t)}{t}. \quad (4)$$

In this work the diffusion constant is calculated in a linear approximation. Secondly we calculate the time-exponent, to see where the behavior of the test particle is no longer Brownian and becomes anomalous. The average behavior of the test particle is obtained by tracking the motion of the particle on the lattice, repeating the calculation 20 times for each case. Based on the attempt rate of the particle, which is the input for a random number generator that gives zero or one, it is determined whether a transition attempt occurs. When a transition attempt occurs, a random direction is chosen. Then it is checked whether this transition can occur, i.e. whether the site is unoccupied, and based on the outcome of this check the transition actually occurs or not.

The average occupation of the lattice is constant during all simulations. However, fluctuations in the occupation of a specific site are incorporated with a certain rate $\Gamma_L$. Using this rate as input for a random number generator producing zero or one, at each time step the sites can switch from being unoccupied to occupied and vice versa. Obviously this means that the case $\Gamma_L = 0$ represents a static lattice. The difference in transition rates of the test particle and the lattice sites will prove to be of great importance in the diffusive behavior of the test particle. Note that the problem formulation makes no energy based assumptions. As mentioned, all states have the same (arbitrary) energy and are thus equivalent. The effects studied in this work have thus a purely kinetic origin.

## III Results

The behavior of the test particle on the disordered square lattice is purely governed by the transition rates (probabilities) and the lattice occupation. As explained in the previous section, if the particle attempts a jump it is in a random direction, and therefore the transition rate should be linearly dependent on the occupation, $\Gamma_{tr} = (1-f)\Gamma_p$, independent on the rate of the lattice. Figure 2 shows the transition rate as a function of lattice occupation for two limit cases: 1) a static lattice, and 2) a lattice that fluctuates with the same rate as the particle attempts a jump. It is found that the transition rate indeed linearly decreases with occupation for both cases. Additionally it is found in a static lattice that the variation in the transition rate becomes significant for $f > 0.5$. The random motion on a static lattice is in principle a percolation problem. Since the site percolation threshold of a square lattice is $f_c = 0.594$ [12], the test particle can be trapped for $f_c > 0.594$, which leads to a larger variation in the transition rate.

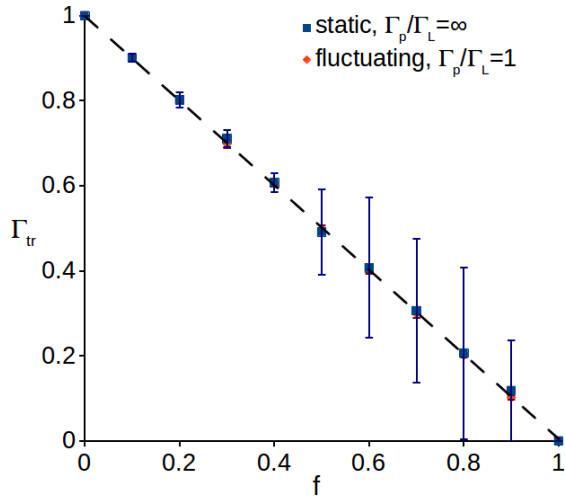

Figure 2: Transition rate as a function of lattice occupation.

The verification that the transition rate is only dependent on the jump attempt rate of the test particle and the occupation of the lattice, verifies that in this formulation the static and fluctuating lattice are energetically equivalent. This also means that if a bi-lattice is constructed in which a half space is fluctuating and the other half space is static, the average time that the test particle spends in both half spaces should be the same. The parameter

$$\phi = \frac{|T_{st} - T_{fl}|}{T_{st} + T_{fl}} \qquad (5)$$

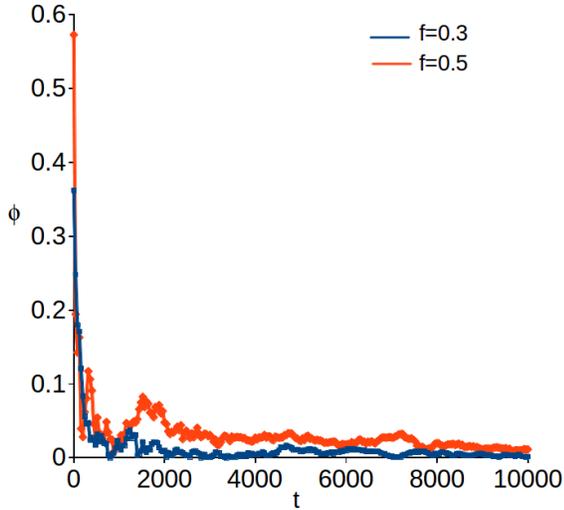

Figure 3: The relative difference between time spend by the test particle in the static and fluctuating part of a bi-lattice with $f = 0.3$ and $f = 0.5$.

where $T_{\text{st}}$ is the total time spend in the static part and $T_{\text{fl}}$ is the total time spend in the fluctuating part of the bi-lattice, should approach zero at large time if the two parts of the bi-lattice are energetically equivalent, and the periodic dimensions of the lattice are chosen small enough for the particle to cover the full periodic space in that time. Of course this analysis can only be performed below the percolation threshold, since above the percolation threshold, not all parts of the static lattice are connected. However, even near the percolation threshold, $f = 0.5$, $\phi$ approaches zero, as shown in Fig. 3. This means that the test particle spends approximately the same amount of time in both the static and fluctuating part of the bi-lattice, again verifying that in this formulation the static and fluctuating lattice are energetically equivalent. This makes the observation in Fig. 4 that the mean square displacement of a test particle (approximately linear with time) on a fluctuating lattice is significantly larger than on a static lattice (both with $f = 0.3$) especially interesting. Figure

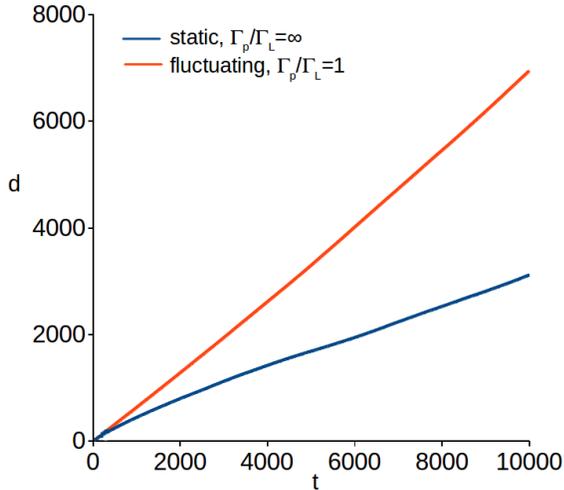

Figure 4: Mean square displacement as a function of time for $f = 0.3$, for a static and fluctuating lattice.

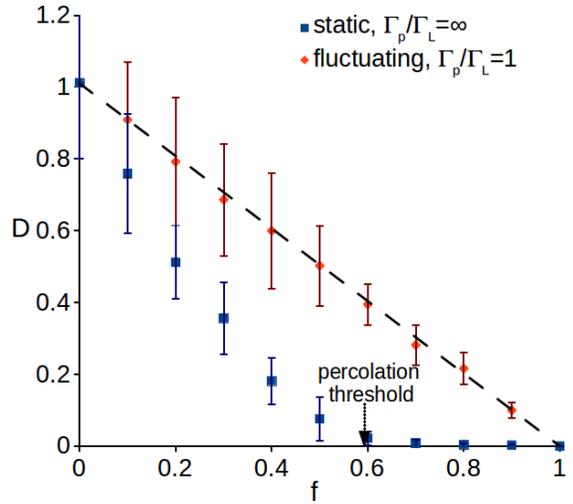

Figure 5: Diffusion constant as a function of lattice occupation.

5 shows the diffusion constant, calculated according to the linear approximation of Eq. 4, as a function of lattice occupation for the two limit cases, the static and fluctuating lattice. For the static lattice the diffusion constant is given by the diffusion constant in a percolation problem on a square lattice known to be $D = D_0(f_c - f)^\mu$, with $1.25 < \mu < 1.33$ [5]. This is indeed found to be true for $D_0 \approx 2$ and $\mu \approx 1.33$. However, diffusion constant of the test particle on a fluctuating lattice is shown to be significantly different. There is no longer a percolation threshold and the diffusion constant decreases linearly with lattice occupation. Due to the fluctuations in the lattice the test particle can make more successive jumps in a particular direction than on a static lattice in the same amount of time.

The diffusive behavior of a particle in a disordered lattice is clearly influenced by the kinetics of the lattice. As seen in Fig. 2, the attempt rate of the particle itself, together with the lattice occupation, governs the actual transition rate. The dependence on the kinetics of the lattice on the diffusion of the particle is therefore best shown by the diffusion constant relative to the transition rate. In Fig. 6, the diffusion constant relative to the transition rate is shown as a function of $\Gamma_p/\Gamma_L$. It is found that for $\Gamma_p/\Gamma_L < 1$, $D/\Gamma_{tr} = 1$, which means that in this regime diffusion is purely controlled by the transition rate of the particle. For $\Gamma_p/\Gamma_L > 1$ the diffusion constant relative to the transition rate decreases, which means that the kinetics of the lattice affects diffusion. For large $\Gamma_p/\Gamma_L$, $D/\Gamma_{tr}$ becomes independent of $\Gamma_p/\Gamma_L$, which represents a static lattice. Figure 6 shows that for larger occupation, the transient regime in $D/\Gamma_{tr}$ is much more steep than for smaller occupation. This means that as occupation increases, the kinetic effects of the lattice on diffusion decreases.

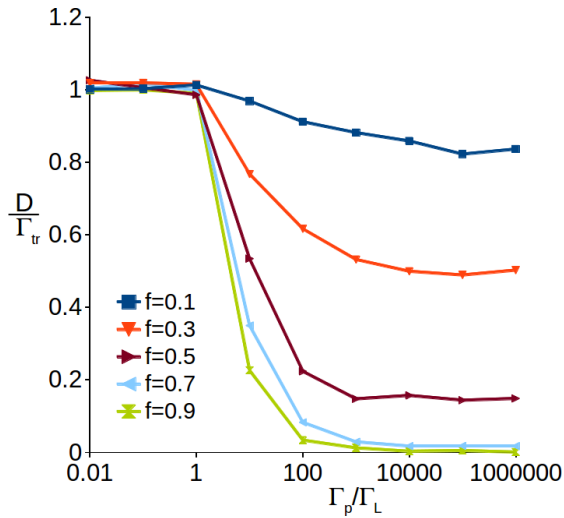

Figure 6: Diffusion constant relative to transition rate as a function of the ratio between attempt rate of the particle and fluctuation rate of the lattice.

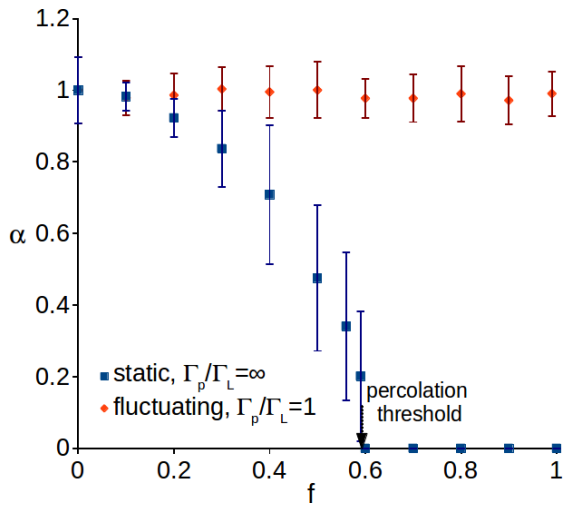

Figure 7: Time-exponent as a function of lattice occupation.

Like mentioned in Section II the mean square displacement is proportional with time by $d(t) \propto t^\alpha$. The diffusion constant in Fig. 5 and 6 is based on a linear approximation which means that the average slope of $d$ versus $t$ gives $D$. However, Fig. 4 shows that the mean square displacement is not always exactly linear with time. Therefore the time-exponent $\alpha$ is calculated for both fluctuating and static lattices and given as a function of occupation in Fig. 7. It is found that $\alpha \approx 1$ for the fluctuating lattice. However, for the static lattice, $\alpha$ decreases and becomes zero at the percolation threshold ($D$ also becomes zero at $f_c$).

This means that in a fast fluctuating lattice the motion of the test particle is Brownian like the particle scatters on other moving particles, while in a static disordered lattice diffusion is anomalous, and becomes more and more sublinear towards the percolation threshold.

The kinetic effects on the measure in which diffusion is anomalous are shown in Fig. 8, where $\alpha$ as a function of $\Gamma_p/\Gamma_L$ is shown. If the lattice fluctuates faster than the particle, the diffusion is Brownian. However, if the lattice fluctuates slower than the particle, the diffusion becomes anomalous, which is more pronounced for larger occupation consistent with Fig. 7.

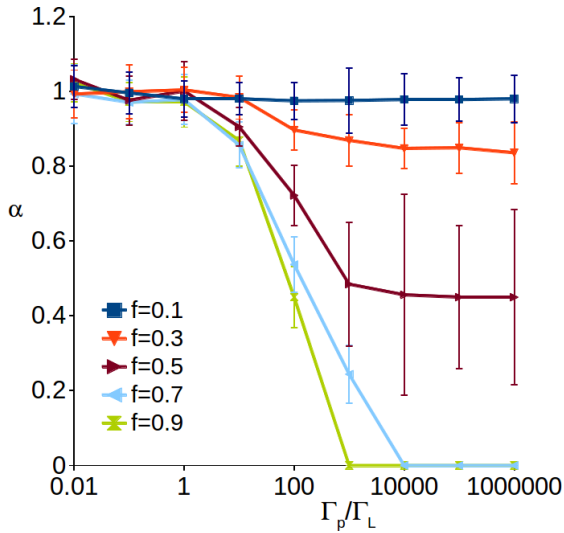

Figure 8: Time-exponent as a function of the ratio between attempt rate of the particle and fluctuation rate of the lattice.

## IV  Conclusion

In this work the kinetic effects of a disordered square lattice on the diffusive motion of a test particle is studied using kinetic Monte Carlo simulations. This leads to the following conclusions.

- Fluctuations in the lattice can kinetically enhance diffusion compared to a static lattice with the same site occupation.

- For $\Gamma_L < \Gamma_p$, diffusion is limited by $\Gamma_L$ and the lattice occupation. However, for $\Gamma_L > \Gamma_p$ the diffusion is limited by $\Gamma_p$ and the lattice occupation.

- Kinetic effects are suppressed by increasing occupation.

- Diffusion is due to Brownian motion on a fast moving lattice, while on a slower moving/static lattice diffusion is anomalous (sublinear).